\documentclass[aps,prl,twocolumn,groupedaddress]{revtex4}

\usepackage{amsmath}

\newcommand*{\Dsl}[0]{{\rlap{\kern2.25pt /}{D}}}
\def\det{{\rm det}}

\newcommand{\be}{\begin{eqnarray}}
\newcommand{\ee}{\end{eqnarray}}
\newcommand\nn{\nonumber}

\begin{document}

\title{The QCD sign problem as a total derivative}

\author{Jeff Greensite$^1$, Joyce C. Myers$^2$ and K. Splittorff$^2$}
\affiliation{$^1$Physics and Astronomy Dept., San Francisco State University, San Francisco, CA 94132, USA \\
$^2$Discovery Center, The Niels Bohr Institute, University of Copenhagen, 
Blegdamsvej 17, DK-2100, Copenhagen {\O}, Denmark}

\date{\today}

\begin{abstract}
We consider the distribution of the complex phase of the fermion 
determinant in QCD at nonzero chemical potential and examine the 
physical conditions under which the distribution takes a Gaussian 
form. We then calculate the baryon number as a function of the 
complex phase of the fermion determinant and show {\sl 1)} 
that the exponential cancellations produced by the sign problem take the form of total derivatives {\sl 2)} that the full baryon number is 
orthogonal to this noise. These insights allow us to define a 
self-consistency requirement for measurements of the baryon 
number in lattice simulations. 
\end{abstract}


\maketitle


{\bf Introduction:} Perhaps the simplest way to identify the asymmetry between matter and anti-matter in QCD is to measure the free energy of a static quark and compare it to that of an anti-quark. These two free energies are linked to the expectation values of an observable, the Polyakov loop, and its complex conjugate. In a baryon asymmetric world the chemical potential, $\mu$, favors propagation of quarks and as a result the expectation value of the Polyakov loop will be different from that of its complex conjugate. If the action was real the two expectation values would be equal. Hence the asymmetry between matter and anti-matter can only occur if the QCD action takes complex values. This fact is known as {\sl the QCD sign problem}, see \cite{deForcrand,Aarts:2013bla} for excellent reviews.

As the physical origin suggests, sign problems likewise occur naturally for systems with strongly correlated electrons \cite{CW}. In both cases the complex nature of the action invalidates the Vafa-Witten theorem \cite{VW}. Vector symmetries such as baryon number and isospin are therefore not protected. In this way the sign problem allows for a much richer phase structure.

The {\sl QCD sign problem} enters through the fermion determinant,  
\begin{equation}
\det(\Dsl+\mu\gamma_0+m) = r e^{i\theta}.
\end{equation}
For $\mu=0$, the determinant is real due to the $\gamma_5$ Hermiticity 
of the Dirac operator (cf.~\cite{Gattringer:2010zz}). The chemical 
potential term breaks $\gamma_5$ Hermiticity and for this reason the
determinant is in general complex (apart from certain gauge groups 
such as $SU(2)$ and $G_2$ with real or pseudo-real representations). 
 While the physical nature and the explicit origin of the sign problem is clear we are only beginning to understand the consequences. The reason is that vacuum expectation values in such non statistical systems have entirely new ways to manifest themselves. The chiral condensate, for example, results from extreme cancellations \cite{OSV,IS}. 

The cancellations due to the sign problem are exponential in the 
volume and thus the use of lattice QCD simulations is severely limited. 
Despite this, remarkable numerical results obtained by 
working at fixed phase of the fermion determinant, and subsequently 
integrating over the distribution of the phase have been presented 
\cite{factorization,FKS,Ejiri}. 
This approach goes by several names: the factorization method 
\cite{factorization}, the density of states method \cite{FKS} 
and the histogram method \cite{Ejiri}.
Here we derive the distribution of the complex phase of the fermion 
determinant at nonzero chemical potential, which is at the core 
of this approach. In particular, we 
examine the conditions under which the distribution 
takes the Gaussian form assumed in \cite{Ejiri}, see \cite{Ejiri-review}
for a review.    

The results presented here follow directly from a general assumption on
the analytic structure of the free energies involved. This assumption is 
identical in nature to that of the replica method, see e.g.~\cite{replica}, and 
hence we expect that the results applies equally generally.

The central challenge to lattice QCD in this context is to compute 
the asymmetry between matter and anti-matter, that is, the average 
baryon number $\langle n_B \rangle_{N_f}$ with $N_f$ dynamical flavors. 
In the approach discussed here one first computes the distribution 
$\langle n_B \delta(\theta - \theta') \rangle_{N_f}$ of the baryon number as 
a function of the complex phase of the fermion determinant. 
(Notation: $\theta$ is the value of the phase which we keep fixed 
and $\theta'$ denotes the value of the phase of the determinant 
for the gauge field configurations in the average.) Then the  
baryon number is the integral of this distribution over $\theta$, 
\be
\langle n_B \rangle_{N_f} = \int {\rm d}\theta ~ \langle n_B \delta(\theta - \theta') \rangle_{N_f} \, .
\ee
In this paper we determine the analytic form of the distribution 
$\langle n_B \delta(\theta - \theta') \rangle_{N_f}$ of the baryon number as 
a function of the complex phase of the fermion determinant. 
The sign problem manifest itself in the complex nature 
of the distribution $\langle n_B \delta(\theta - \theta') \rangle_{N_f}$. 
In particular, as we will demonstrate below, terms which are 
sub-leading in the volume at fixed $\theta$ contribute at 
leading order after integration over $\theta$. This 
makes the $\theta$-integral hard to control numerically.   
The analytic form of the distribution of the baryon number as a function 
of the complex phase, however, reveals two important properties: 
{\sl first}, the cancellations produced by the sign problem 
takes the form of total derivatives wrt.~$\theta$,
 {\sl second}, the full baryon number (the signal) is orthogonal to this noise. 
Based on this we formulate a self-consistency requirement 
for measurements of the baryon number. 

The general assumption and results obtained are exemplified within 
the hadron resonance gas model, and in the combined strong coupling 
and hopping parameter expansion. 

\vspace{3mm}

{\bf Distribution of the phase:} We first consider 
the distribution, $\langle \delta (\theta - \theta') \rangle_{N_f}$, 
of the complex phase. Using the basic definitions we have \cite{SV}
\be
&& \langle \delta (\theta - \theta') \rangle_{N_f} \nn\\
&= &\frac{1}{Z_{N_f}} \int {\rm {\cal D}}A ~ \delta(\theta - \theta') {\det}^{N_f} (\Dsl+\gamma_0 \mu+m) e^{-S_{YM}}\nn\\
&= &\frac{1}{Z_{N_f}} \int {\rm {\cal D}}A ~ \delta(\theta - \theta') \left| \det (\Dsl+\gamma_0 \mu+m) \right|^{N_f} e^{N_f i \theta'(A)}\nn\\
&&\hspace{2mm}\times e^{-S_{YM}} \nn\\
&= &e^{N_f i \theta} ~ \frac{Z_{|N_f|}}{Z_{N_f}} ~ \langle \delta(\theta-\theta') \rangle_{|N_f|}.\label{gen-th-dist} 
\ee
The subscript $|N_f|$ refers to the phase quenched theory where the fermion determinant is replaced by its absolute value. The phase quenched distribution, $\langle \delta(\theta-\theta') \rangle_{|N_f|}$, is of course real and positive. The distribution in the full theory, however, has an explicit phase factor $\exp(N_f i \theta)$. This explicit phase must give rise to exponential cancellations in order to fulfill the normalization
\be
\int {\rm d}\theta ~ \langle \delta (\theta - \theta') \rangle_{N_f} = 1 \, 
\ee
because $Z_{|N_f|}/Z_{N_f}$ is exponentially large in the volume.

To compute the distribution of the complex phase analytically we express it through a Fourier transform \cite{LSV}
\be
\langle \delta(\theta - \theta') \rangle_{N_f} = \frac{1}{2\pi}\sum_{p=-\infty}^{\infty} e^{-i p \theta} \langle e^{i p \theta'} \rangle_{N_f} \, ,
\label{F-trans}
\ee
where the moments are
\be
\langle e^{i p \theta'} \rangle_{N_f} =
\frac{1}{Z_{N_f}} \bigg{\langle} \frac{\det^{N_f+p/2} (\Dsl+\gamma_0 \mu + m)}{\det^{p/2}(\Dsl-\gamma_0 \mu + m)} \bigg{\rangle} \, ,
\label{complex-phase-moms}
\ee
and the bracket without a subscript is the average w.r.t.~the 
Yang-Mills action. 
When $p$ is even, the moments are simply partition functions with $p/2$ additional quarks and $p/2$ additional quarks of bosonic statistics (the complex conjugate of the determinant is equivalent to a sign change of the chemical potential). However, the expression (\ref{complex-phase-moms}) is perfectly well-defined also when $p$ is odd.

It is natural to assume that the associated free energies have a series 
expansion in $p$, ie.~that $\log\langle e^{i p \theta'} \rangle$ 
is a polynomial in $p$.
The form of this polynomial is constrained by the following requirements: \\

{\sl 1)} Since partition functions are even in $\mu$ the 
moments, Eq.~(\ref{complex-phase-moms}), are invariant under 
$p/2\to-p/2-N_f$ \\

 {\sl 2)} If we take 
$p/2\to p/2-N_f/2$ the moments become proportional to the phase quenched moments, 
\be
\langle e^{i (p-N_f) \theta'} \rangle_{N_f} = \frac{Z_{|N_f|}}{Z_{N_f}}\langle e^{i p \theta'} \rangle_{|N_f|}.
\ee
To ensure these requirements the unquenched moments take the form
\be
&& \langle e^{i p \theta'} \rangle_{N_f} \\
& = & e^{-p/2(p/2+N_f)X_1-(p/2(p/2+N_f))^2X_2-(p/2(p/2+N_f))^3X_3-\ldots}. \nn
\label{mom}
\ee
The values of $X_j$ determine the distribution of the complex phase 
through Eq.~(\ref{F-trans}). A few properties of the $X_j$ can be deduced 
from first principles: First of all, 
since the moments are partition functions 
the coefficients $X_j$ are extensive. Moreover, as noted above we have 
\be
\label{mom-rel}
 & & \langle e^{i p \theta'} \rangle_{|N_f|} \\
 & = & \frac{Z_{N_f}}{Z_{|N_f|}}e^{-(p^2-N_f^2)X_1/2^2-(p^2-N_f^2)^2X_2/2^4-(p^2-N_f^2)^3X_3/2^6\ldots}, \nn
\ee
and in particular for $p=0$
\be
1 = \frac{Z_{N_f}}{Z_{|N_f|}}e^{(N_f/2)^2X_1-(N_f/2)^4X_2+(N_f/2)^6X_3\ldots}.
\label{mom-rel-peq0}
\ee
The sum in the exponent is therefore the free energy difference between the 
full and the phase quenched theory and hence must be positive since 
$Z_{N_f}\leq Z_{|N_f|}$. 

\vspace{3mm}

{\bf Gaussian distribution:} In \cite{Ejiri} it was found numerically that 
the distribution of the phase to good approximation assumes a Gaussian form. 
To obtain the Gaussian form of 
$\langle \delta (\theta'-\theta) \rangle_{|N_f|}$ requires that only 
$X_1$ is nonzero, ie.~that $X_j=0$ for all $j>1$. In that case we
get 
\be
\langle \delta(\theta - \theta') \rangle_{N_f} = \frac{e^{i N_f\theta}}{\sqrt{\pi X_1}}e^{(N_f/2)^2X_1}\sum_{k=-\infty}^\infty e^{-(\theta+2\pi k)^2/X_1}.
\label{Gaussian}
\ee 
The sum is a compactified Gaussian with a width that grows as the 
square root of the volume (recall that $X_1$ is extensive).
Note that this fits perfectly with the general relation, 
Eq.~(\ref{gen-th-dist}), using Eq.~(\ref{mom-rel-peq0}). The Gaussian 
form is the key approximation in the approach of \cite{Ejiri,Ejiri-review}. 
Therefore it is desirable to understand under which conditions we can 
expect a Gaussian distribution of the phase.

To examine if conditions exist such that $X_j=0$ for all $j>1$, 
analytically from first principles is at present not feasible since it 
is clearly as hard as to evaluate the full QCD partition function. 
However, from the form of the moments (\ref{mom}) it is clear 
that any Feynman diagram with more than three quark lines which contributes 
to the free energy of the moments will lead to a non-Gaussian form. 
Let us exemplify this general analytic understanding of the conditions 
under which the Gaussian follows by two examples.

{\sl First}, suppose we evaluate the moments 
Eq.~(\ref{complex-phase-moms}) using
the hadron resonance gas model. This involves a 1-loop 
computation of all possible mesons and baryons which 
can be formed from the $p+N_f$ quark flavors (In practice 
these states correspond to the decomposition of 
${\bf n} \otimes {\bar {\bf n}}$ and 
${\bf n} \otimes {\bf n} \otimes {\bf n}$ in $SU(2(p+N_f))$ to $SU(p+N_f)_{flavor} \times SU(2)_{spin}$, followed by collection of weights with each spin degeneracy. All details will be provided in 
\cite{GMS}.). Since all possible meson and baryon states have 
either 2 or 3 internal quark lines the combinatorics at 1-loop is 
unable to generate terms proportional to $(p/2(p/2+N_f))^2$ or higher.
Hence, the summation upon $p$ in Eq.~(\ref{F-trans}) necessarily leads 
to the Gaussian form in the hadron resonance gas as long as we work at 
1-loop order.

{\sl Second}, if we instead compute the moments Eq.~(\ref{complex-phase-moms}) 
in the combined strong coupling and hopping parameter expansion, then to 
third order in the hopping parameter 
(see \cite{Owe} for the form of the hopping parameter expansion 
to third order) it is again impossible to generate $(p/2(p/2+N_f))^2$ and 
higher order terms, simply because the number of winding loops is 
insufficient. Explicitly in the confined phase, we find order by order in the 
hopping parameter $h$ (see \cite{GMS} for details), 
\be
X_1 = 4 a_1^2h^2 \sinh^2(\mu/T) N_s  \ \ {\rm and} \ \  X_{j>1}=0   
\label{X1strong}
\ee
to second order,
\be
X_2 = - \frac{4}{3}a_1^4 h^4 \sinh^4(\mu/T) 18 \lambda_1 N_s \ \  {\rm and} \ \ X_{j>2}=0
\label{X2strong}
\ee
to 4th order and finally 
\be
X_3 & = &- \frac{8}{45} a_1^6h^6 \sinh^6(\mu/T)  (5 + 450 \lambda_1) N_s \\
&& \hspace{2cm} \ \ {\rm and} \ \ X_{j>3}=0 \nn  
\label{X3strong}
\ee
to 6th order. $N_s$ is the lattice volume and thus the $X_j$ are extensive as
expected. 
The coupling $\lambda_1$ is the leading coupling in an effective
Polyakov line action derived from the Wilson action via a strong 
coupling expansion. Its precise relation to the gauge coupling is given 
in ref.~\cite{Owe}. The constant $a_1$, which is ${\cal O}(1)$, 
depends on the choice of lattice action. 

We see that, 
starting at fourth order in the hopping parameter expansion terms of order 
$(p/2(p/2+N_f))^2$ appear naturally.
Furthermore, from Eqs.~(\ref{X1strong})-(\ref{X3strong}) we observe 
that the corrections to the Gaussian form set in at increasing order 
in $\mu/T$. This is completely general: If we Taylor expand the free 
energy of the moments Eq.~(\ref{complex-phase-moms}) in $\mu/T$, 
terms of order $(p/2(p/2+N_f))^j$ are present at order 
$(\mu/T)^{2j}$ and higher. To 
order $(\mu/T)^{2j}$ the terms of order $(p/2(p/2+N_f))^j$ are proportional to 
the $2j$th order cumulant of $\theta'$ considered in \cite{Ejiri-cumulant}.

We conclude that one generically finds non-Gaussian corrections 
to the distribution of the phase in QCD. However,
as we have seen, the Gaussian approximation has a natural interpretation 
in terms of combinatorical factors resulting from quark lines. If the combinatorics in evaluating the moments of the phase factor 
involve at most three quark lines then the Gaussian form results.

As we shall now see, the distribution of the baryon number over $\theta$ 
helps us to understand what it takes to make the Gaussian approximation 
self-consistent for measurement of the baryon number.

\vspace{3mm}

{\bf Distribution of the baryon number:} To access the baryon number 
distribution we again proceed via the Fourier transform
\be
\langle n_{B} \delta(\theta - \theta') \rangle_{N_f} = \frac{1}{2 \pi} \sum_{p=-\infty}^{\infty}  e^{-i p \theta} \langle n_{B} e^{i p \theta'} \rangle_{N_f} \, ,
\label{baryon-num-dist}
\ee
where the moments are given by
\be
&&\langle n_{B} e^{i p \theta'} \rangle_{N_f} = \frac{1}{Z_{N_f}}\\
&& \times \lim_{{\tilde \mu} \rightarrow \mu} \frac{\partial}{\partial {\tilde \mu}} \bigg{\langle} \frac{\det^{p/2} (\Dsl + \gamma_0 \mu + m)}{\det^{p/2} (\Dsl - \gamma_0 \mu + m)} {\det}^{N_f} (\Dsl + \gamma_0 {\tilde \mu} + m) \bigg{\rangle} . \nn
\label{baryon-num-moms} 
\ee
The $p$-dependence of the log of these moments will of course reduce to a polynomial in 
$p/2(p/2+N_f)$ at $\tilde\mu=\mu$, however, before we take this limit we must 
differentiate wrt $\tilde\mu$. For $\tilde\mu\neq\mu$ the $p$-dependence 
of free energy in the moments can be a more general polynomial in $p$
\be
& &\frac{1}{Z_{N_f}}\bigg{\langle} \frac{\det^{p/2} (\Dsl + \gamma_0 \mu + m)}{\det^{p/2} (\Dsl - \gamma_0 \mu + m)} {\det}^{N_f} (\Dsl + \gamma_0 {\tilde \mu} + m) \bigg{\rangle}\nonumber\\ 
& = & \exp(k_0+k_1p+k_2p^2+\ldots),
\label{baryon-num-moms}
\ee
for suitable functions $k_i$. This directly leads to 
\be
\label{baryon-num-moms-gen-result}
 && \langle n_{B} e^{i p \theta'} \rangle_{N_f}      \\
 & = & (c_0 + c_1 p + c_2 p^2+\ldots) e^{-p/2(p/2+N_f)X_1+\ldots} , \nn
\ee
where $c_j\equiv\lim_{\tilde\mu\to\mu}\partial_{\tilde \mu}k_j$. We conclude that 
the distribution of the baryon number takes the form
\be\label{nBdist}
\langle n_{B} \delta(\theta - \theta') \rangle_{N_f}  
&=& \frac{1}{2\pi} \sum_{p=-\infty}^{\infty} e^{-i p \theta}  \\
&& \hspace{-19mm}\times\left( c_0 + c_1 p + c_2 p^2 +\ldots\right) e^{-p/2 (p/2+N_f)X_1+\ldots} .\nn
\ee
This expression looks rather complicated, however, 
as we shall now demonstrate the entire contribution to the average baryon 
number comes from the $c_0$ term. The remainder of the terms are noise. 
\vspace{3mm}

{\bf Total derivative:} To see how the integration over $\theta$ singles 
out the $c_0$ term we now 
rewrite the distribution of the baryon number in terms of total derivatives 
of the distribution of the phase itself. One easily verifies from Eq.~(\ref{nBdist}) that 
\be
\label{nBdist-totderiv}
&&\langle n_{B} \delta(\theta - \theta') \rangle_{N_f}\\
&=& \frac{1}{2\pi} \sum_{p=-\infty}^{\infty}  \left( c_0 + \frac{c_1}{-i} \frac{\partial}{\partial \theta} + \frac{c_2}{(-i)^2} \frac{\partial^2}{\partial \theta^2}+\ldots \right) \nn\\
&& \hspace{2cm} \times e^{-i p \theta} e^{-p/2 (p/2+N_f)X_1+\ldots} \nn \\
&=& \left( c_0 + \frac{c_1}{-i} \frac{\partial}{\partial \theta} + \frac{c_2}{(-i)^2} \frac{\partial^2}{\partial \theta^2}+\ldots \right) 
\langle \delta(\theta - \theta') \rangle_{N_f} \nn
\, .
\ee 
This amazing relation is completely general, our only assumption is that 
the free energy of the moments are polynomials in $p$. 
The integration over $\theta$ needed to obtain the full baryon number is now 
trivial 
\be
\langle n_{B} \rangle_{N_f} = \int d\theta \ \langle n_{B} \delta(\theta - \theta') \rangle_{N_f} = c_0 . 
\label{th-int-nB}
\ee
The terms proportional to $c_i$ with $i\geq1$ hence constitute the 
background noise within which the baryon number signal $c_0$ is to be measured.
As we will demonstrate below it is essential 
to exclude the noise in order to obtain the average baryon number.  

\vspace{3mm}

{\bf Baryon number in the Gaussian approximation:} 
To see that it is essential to exclude the noise let us again consider 
the special case where $X_j=0$ for $j>1$ 
and the distribution of the phase is given by the Gaussian form 
(\ref{Gaussian}).
To get the baryon number distribution we take derivatives 
of the Gaussian, cf.~Eq.~(\ref{nBdist-totderiv}).  
Explicitly we get \footnote{In \cite{LSV} the distribution 
of the phase and of the baryon number as a function of  
the phase was computed within 1-loop chiral perturbation 
theory. Since there are only two quark lines in the loop the computation  
necessarily resulted in a Gaussian distribution. Moreover,
since pions are baryon number neutral only the $c_1$ term resulted. The 
1-loop computation within chiral perturbation theory therefore did not
reveal the 
general way that a nonzero baryon number forms.}
\be\label{baryon-num-dist-result}
\langle n_{B} \delta(\theta - \theta') \rangle_{N_f} & = & \frac{1}{\sqrt{\pi X_1}}e^{N_f^2 X_1/4 + N_f i \theta}  \\
&& \hspace{-30mm}\times\sum_{k=-\infty}^\infty e^{-(\theta+2\pi k)^2/X_1}\big[ c_0 - c_1 (N_f + i \frac{2(\theta+2\pi k)}{X_1}) \nn \\
&& \hspace{-14mm}+ c_2 \big((N_f + i \frac{2(\theta+2\pi k)}{X_1})^2 + \frac{2}{X_1}\big) +\ldots\big] . \nn
\ee
The total derivatives of Eq.~(\ref{nBdist-totderiv}) include terms of 
different powers of volume (recall that $X_1$ is extensive). 
This is {\sl the core of the sign problem for measurements of the baryon number}: Terms which are down 
with powers of volume in the distribution contribute at leading order to 
the baryon number. If the contributions from (\ref{baryon-num-dist-result}) 
that go like $1/X_1$, $1/X_1^2$, ... are dropped the result is
$c_0-c_1N_f+c_2N_f^2+\ldots$ rather than $c_0$ as required, 
cf.~Eq.~(\ref{th-int-nB}).

\vspace{3mm}

{\bf Self-consistency requirement:}
As we have just seen it is essential to control the noise terms in 
order to measure the baryon number. In other words we must measure 
the distribution of the phase and check that we can control the error
on the derivatives of the distribution. Clearly if we would 
use brute force to measure all derivatives the numerical problem would be 
exponentially hard.

However, let us again turn to the analytic cases of the 1-loop hadron 
resonance gas and the strong coupling expansion to third order in the hopping 
parameter, where the distribution of the phase automatically becomes 
a Gaussian: in both cases the polynomial in the exponent of 
Eq.~(\ref{baryon-num-moms}) is at most third order in $p$ and 
no higher than a second order derivative can appear 
in Eq.~(\ref{nBdist-totderiv}).   
Therefore, to make the Gaussian approximation self-consistent for the 
measurement of the baryon number it is sufficient 
to measure the distribution of $\theta$ to an accuracy which is quadratic 
in the inverse volume. While this is a demanding numerical task it is 
clearly advantageous compared to exponential accuracy.   

Similarly, it is possible to consistently include corrections to the 
Gaussian coming from $X_j$ with $j>1$ up to a given order $j=J$ if we are able 
to control the errors on the derivatives up to order $2J$. 

\vspace{3mm}

{\bf Orthogonality of signal and noise:} Such higher 
order terms in $\theta$ vary increasingly rapidly with $\theta$ and thus 
are harder to capture. 
The Gaussian example, however, suggests a possible way out of this.
Let us rewrite the distribution of the baryon number in 
Eq.~(\ref{baryon-num-dist-result}) as 
\be
\langle n_B\delta(\theta-\theta') \rangle_{N_f} 
&= &\frac{1}{\sqrt{\pi X_1}} \sum_{k=-\infty}^{\infty} \bigg[ c_0 {\rm H}_0\left(\tfrac{i X_1 N_f/2 - \theta-2\pi k}{\sqrt{X_1}}\right)  \nn \\
&& \hspace{-7mm}+ \frac{i c_1}{\sqrt{X_1}} {\rm H}_1\left(\tfrac{i X_1 N_f/2 - \theta-2\pi k}{\sqrt{X_1}}\right)+\ldots \bigg] \nonumber \\
&& \times e^{-(i X_1 N_f/2 - \theta-2\pi k)^2/X_1}, 
\label{n_B}
\ee
where $H_n$ are the Hermite polynomials (It is of course not accidental 
that the Hermite polynomials appear, it follows because the Gaussian is the 
generating function and we take derivatives, cf.~Eq.~(\ref{nBdist-totderiv}).). 
After a shift of the contour 
by $iX_1N_f/2$ the integral over $\theta$ in (\ref{th-int-nB}) follows 
directly from the orthogonality of the Hermite polynomials 
\be
&& \int {\rm d}x ~ {\rm H}_m(x) {\rm H}_n(x) e^{-x^2} = 2^n n ! \sqrt{\pi} \delta_{m n} \ ,
\ee
and the fact that $H_0=1$. Note that the shift along the imaginary axis is 
proportional to the volume.   
Moreover, due to the orthogonality of the Hermite polynomials on the 
Gaussian weight, the signal $c_0H_0$ is in this sense orthogonal to 
the noise produced by $\sum_{i\geq 1}c_iH_i$.
 
The orthogonality is also present in general where the noise  
is orthogonal to the signal in the following sense
\be
&& \hspace{-4mm} \int d\theta \frac{\langle n_{B} \delta(\theta - \theta') \rangle_{\rm c_0}}{\langle \delta(\theta - \theta') \rangle_{N_f}} \frac{\langle n_{B} \delta(\theta - \theta') \rangle_{\rm c_{j>0}}}{\langle \delta(\theta - \theta') \rangle_{N_f}} \langle \delta(\theta - \theta') \rangle_{N_f} \nn \\
& & =  0 . 
\ee
This follows directly from the general relation, Eq.~(\ref{nBdist-totderiv}), 
where the signal is given by the $c_0$ term and the noise is the remainder, 
ie.~the terms with $j>0$. Note that since the pions do not carry baryon 
number they only contribute to the $c_j$ with $j>0$, see \cite{LSV,GMS} 
for details.

As demonstrated for unitary fermions \cite{David-plenary} analytic 
insights on the form of the noise can be turned into a practical numerical 
tool. If the orthogonality found above can be implemented numerically it 
has the potential to cancel the entire pion noise from the measurement of 
the baryon number.

\vspace{2mm}

{\bf Conclusions:} The analysis of the moments has given us a physical 
understanding of the Gaussian approximation which is central to the 
histogram method: if the combinatorics in evaluating the moments of 
the phase factor involve at most
three quark lines then the Gaussian form results. The Gaussian 
approximation therefore potentially 
captures the dynamics of the 1-loop hadron resonance gas model as well 
as that of the strong coupling expansion to third order in the hopping 
parameter. 
However, we also note that non-Gaussian terms are generic, and 
increase in importance as the hopping parameter, chemical potential 
$\mu/T$ and lattice coupling $\beta$ increase. These terms may therefore be 
quite significant for real QCD, which of course entails light
quark masses and the $\beta \to \infty$ limit.

The results of this paper follow from the assumption that the free 
energy of the moments can be written as a polynomial in $p$. Since 
this is equivalent in nature to the central assumption of the 
replica method, see e.g.~\cite{replica}, we expect that it applies equally 
generally. 
Finally, when $\mu > \frac{m_{\pi}}{2}$ where 
$m_\pi$ is the mass of the pion the moments are dominated by 
Bose-Einstein condensation of pions.  
Such a condensate gives rise to a linear term in $p$, see \cite{LSV}.
It would be most interesting to extend the results of this paper also
to this region.

\vspace{3mm}

{\bf Acknowledgments:} 
We would like to thank Jan Rosseel, Simon Hands, Tim Hollowood, Gert Aarts, 
as well as organizers and participants of the INT program on 'Quantum Noise' 
for discussions. This work was supported 
by U.S. DOE Grant No.~DEFG03-92ER40711 (JG) and the {\sl Sapere Aude} program 
of The Danish Council for Independent Research (JCM and KS).

\end{document}